\documentclass[aps,twocolumn,noshowkeys,superscriptaddress,noshowpacs]{revtex4}

\usepackage{dcolumn}
\usepackage{amsmath}
\usepackage{bm}
\usepackage{nicefrac}
\usepackage{graphicx}
\usepackage{epsfig}
\usepackage{color}
\usepackage{epstopdf}

\begin{document}

\title{Spin Pumping from Permalloy into Uncompensated Antiferromagnetic Co doped Zinc Oxide}
\date{\today}

\author{Martin Buchner}
\email{martin.buchner@jku.at; Phone: +43-732-2468-9651; FAX: -9696}
\affiliation{Institut f{\"u}r Halbleiter- und Fest\-k{\"o}r\-per\-phy\-sik, Johannes Kepler Universit{\"a}t, Altenberger Str. 69, 4040 Linz, Austria}
\author{Julia Lumetzberger}
\affiliation{Institut f{\"u}r Halbleiter- und Fest\-k{\"o}r\-per\-phy\-sik, Johannes Kepler Universit{\"a}t, Altenberger Str. 69, 4040 Linz, Austria}
\author{Verena Ney}
\affiliation{Institut f{\"u}r Halbleiter- und Fest\-k{\"o}r\-per\-phy\-sik, Johannes Kepler Universit{\"a}t, Altenberger Str. 69, 4040 Linz, Austria}
\author{Tadd{\"a}us Schaffers}
\altaffiliation{Current address: NanoSpin, Department of Applied Physics, Aalto University School of Science, P.O. Box 15100, FI-00076 Aalto, Finland}
\affiliation{Institut f{\"u}r Halbleiter- und Fest\-k{\"o}r\-per\-phy\-sik, Johannes Kepler Universit{\"a}t, Altenberger Str. 69, 4040 Linz, Austria}
\author{Ni{\'e}li Daff{\'e}}
\affiliation{Swiss Light Source, Paul Scherrer Institut, CH-5232 Villigen PSI, Switzerland}
%\author{Cinthia Piamonteze}
%\affiliation{Swiss Light Source, Paul Scherrer Institut, CH-5232 Villigen PSI, Switzerland}
\author{Andreas Ney}  
\affiliation{Institut f{\"u}r Halbleiter- und Fest\-k{\"o}r\-per\-phy\-sik, Johannes Kepler Universit{\"a}t, Altenberger Str. 69, 4040 Linz, Austria}

\begin{abstract}
Heterostructures of Co-doped ZnO and Permalloy were investigated for their static and dynamic magnetic interaction. The highly Co-doped ZnO is paramagentic at room temperature and becomes an uncompensated antiferromagnet at low temperatures, showing a narrowly opened hysteresis and a vertical exchange bias shift even in the absence of any ferromagnetic layer. At low temperatures in combination with Permalloy an exchange bias is found causing a horizontal as well as vertical shift of the hysteresis of the heterostructure together with an increase in coercive field. Furthermore, an increase in the Gilbert damping parameter at room temperature was found by multifrequency FMR evidencing spin pumping. Temperature dependent FMR shows a maximum in magnetic damping close to the magnetic phase transition. These measurements also evidence the exchange bias interaction of Permalloy and long-range ordered Co-O-Co structures in ZnO, that are barely detectable by SQUID due to the shorter probing times in FMR. 
\end{abstract} 

\maketitle

\section*{I. Introduction}

In spintronics a variety of concepts have been developed over the past years to generate and manipulate spin currents \cite{ZFS04,SZ12}. Amongst them are the spin Hall effect (SHE), which originates from the spin orbit coupling \cite{DP71}, spin caloritronics \cite{BSW12} utilizing the spin seebeck effect \cite{UTH08} or spin transfer torque (current induced torque) due to angular momentum conservation \cite{BKO12} as examples. Spin pumping \cite{TBB02}, where a precessing magnetization transfers angular momentum to an adjacent layer, proved to be a very versatile method since it has been reported for different types of magnetic orders \cite{MAM01, TYI08, IAS11, MGL14} or electrical properties \cite{ATI11, ZJJ14, HLN14} of materials. Furthermore it could also be verified in trilayer systems where the precessing ferromagnet and the spin sink, into which the angular momentum is transferred, are separated by a non-magnetic spacer \cite{LTL04, MKB12, KSA13, BFP16}. This is strongly dependent on the material, while for Cu \cite{LTL04}, Au \cite{MKB12}, or Al \cite{KSA13} pumping through a few nanometers is possible an MgO barrier of 1\,nm is enough to completely suppress spin pumping \cite{BFP16}.

Spintronic devices are usually based on a ferromagnet (FM) although antiferromagnetic spintronics \cite{BMT18} holds the advantages of faster dynamics, less perturbation by external magnetic fields and no stray fields. The latter two are caused by the zero net magnetization of an antiferromagnet (AFM), which on the other hand makes them harder to manipulate. One way to control an AFM is by using an adjacent FM layer and exploiting the exchange-bias (EB) effect \cite{MB56, NS99}. Measuring spin-transfer torque in FM/AFM bilayer structures, is possible \cite{WSN07, WBA09}, but challenging due to Joule heating \cite{TZS07, DTH08, TZS10} or possible unstable antiferromagnetic orders \cite{UA07}. Antiferromagnets can be used either as spin source \cite{THY14} or as spin sink \cite{MGL14, FOA16} in a spin pumping experiment. Thereby the spin mixing conductance, a measure for the absorption of angular (spin) momentum at the interface \cite{TBB02}, is described by intersublattice scattering at an antiferromagnetic interface \cite{HM08}. Linear response theory predicted an enhancement of spin pumping near magnetic phase transitions \cite{OAS14}, which could recently also be verified experimentally \cite{FOA16}.

In this work we investigate the behavior of the uncompensated, antiferromagnetic Co$_\text{x}$Zn$_\text{1-x}$O with x $\in$ \{0.3, 0.5, 0.6\} (in the following 30\,\%, 50\,\% and 60\,\% Co:ZnO) in contact to ferromagnetic permalloy (Py). While weakly paramagnetic at room temperature, Co:ZnO makes a phase transition to an antiferromagnetic state at a N\'{e}el temperature ($T_\text{N}$) dependent on the Co concentration \cite{NHL16}. This resulting antiferromagnetism is not fully compensated which is evidenced by a narrow hysteresis and a non saturating magnetization up to 17\,T \cite{HNO15}. Furthermore, Co:ZnO films exhibit a vertical EB in complete absence of a FM layer \cite{HNS16}. This vertical exchange shift is dependent on the Co concentration \cite{NHL16}, temperature and cooling field \cite{BHN19} and the field imprinted magnetization predominantly shows orbital character \cite{BHN18}. Note that below the coalesence limit of 20\,\% the vertical EB vanishes. Co:ZnO therefore offers to study magnetic interactions between an uncompensated AFM and a FM Py layer. Static coupling, visible as EB, is investigated using super conducting quantum interference device (SQUID) magnetometry. The dynamic coupling across the interface is measured using ferromagnetic resonance (FMR) at room temperature and around the magnetic transition temperatures determined from $M(T)$ SQUID measurements. Element selective XMCD studies are carried out to disentangle the individual magnetic contributions. Finally heterostructures with an Al spacer were investigated to rule out intermixing at the interface as source for the coupling effect.

\section*{II. Experimental Details}

Heterostructures consisting of Co:ZnO, Py and Al, as shown in Fig.\,\ref{figure1} were fabricated on c-plane sapphire substrates using reactive magnetron sputtering (RMS) and pulsed laser deposition (PLD) at a process pressure of 4 $\times$ 10$^{\text{-3}}$\,mbar. The different layers of a heterostructure are all grown in the same UHV chamber with a base pressure of 2 $\times$ 10$^{\text{-9}}$\,mbar in order to ensure an uncontaminated interface. While Py and Co:ZnO are grown by magnetron sputtering, the Al spacer and capping layers are grown by PLD. Al and Py are fabricated at room temperature using 10 standard cubic centimeters per minute (sccm) Ar as a process gas. 

For the heterostructures containing a Co:ZnO layer, samples with three different Co concentrations of 30\,\%, 50\,\% and 60\,\% are grown utilizing preparation conditions that yield the best crystalline quality known for Co:ZnO single layers \cite{HNO15,NHL16, BHN18}. For 30\,\% and 50\,\% Co:ZnO metallic sputter targets of Co and Zn are used at an Ar:O$_\text{2}$ ratio of 10\,:\,1 sccm, while for 60\,\% Co:ZnO no oxygen and a ceramic composite target of ZnO and Co$_\text{3}$O$_\text{4}$ with a 3\,:\,2 ratio is used. The optimized growth temperatures are 450\,$^\circ$C, 294\,$^\circ$C and 525\,$^\circ$C. Between Co:ZnO growth and the next layer a cool-down period is required, to minimize inter-diffusion between Py and Co:ZnO. 

The static magnetic properties are investigated by SQUID magnetometry. $M(H)$ curves are recorded at 300\,K and 2\,K in in-plane geometry with a maximum magnetic field of $\pm$\,5\,T. During cool-down either a magnetic field of $\pm$\,5\,T or zero magnetic field is applied to differentiate between plus-field-cooled (pFC), minus-field-cooled (mFC) or zero-field-cooled (ZFC) measurements. All measurements shown in this work have been corrected by the diamagnetic background of the sapphire substrate and care was taken to avoid well-known artifacts \cite{SSN11, BHH18}.

For probing the element selective magnetic properties X-ray absorption (XAS) measurements were conducted at the XTreme beamline \cite{PFS12} at the Swiss Synchrotron Lightsource (SLS). From the XAS the X-ray magnetic circular dichroism (XMCD) is obtained by taking the direct difference between XAS with left and right circular polarization. The measurements were conducted with total fluoresence yield under 20$^\circ$ grazing incidence. Thereby, the maximum magnetic field of 6.8\,T was applied. Both, external magnetic field and photon helicity have been reversed to minimize measurement artefacts. Again pFC, mFC and ZFC measurements were conducted applying either zero or the maximum field in the respective direction. 

The dynamic magnetic properties were measured using multi-frequency and temperature dependent FMR. Multi-frequency FMR is exclusively measured at room temperature from 3\,GHz to 10\,GHz using a short circuited semi-rigid cable \cite{RMW12}. Temperature dependent measurements are conducted using an X-band resonator at 9.5\,GHz. Starting at 4\,K the temperature is increased to 50\,K in order to be above the N$\acute{\text{e}}$el-temperature of the Co:ZnO samples \cite{NHL16, BHN19}. At both FMR setups the measurements were done in in-plane direction. 

The measured raw data for SQUID, FMR, XAS and XMCD can be found in a following data repository \cite{BLN19}.

\section*{III. Experimental results \& Discussion}

\begin{figure}[h]
	\centering
		\includegraphics[width=0.35\textwidth]{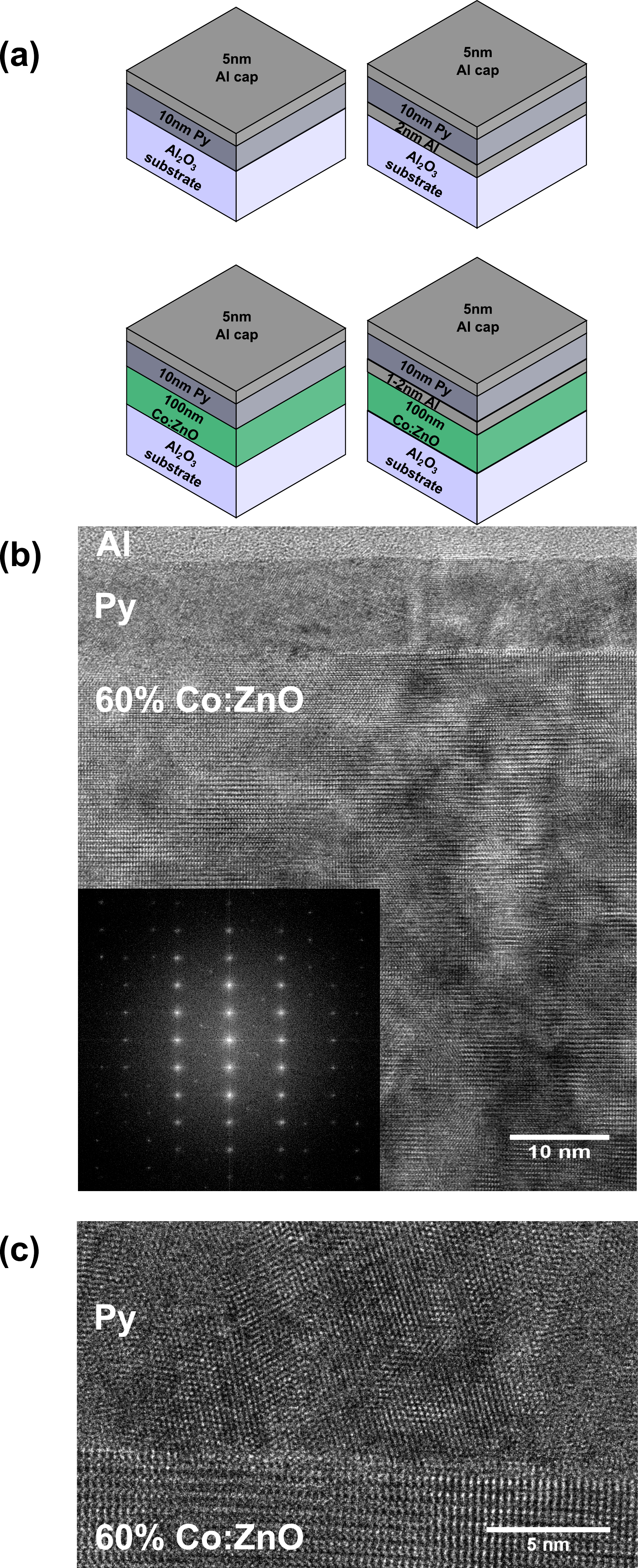}
		\vspace{-0.2cm}
	\caption{(a) shows the schematic setup of the samples. For the Co:ZnO layer three different Co concentrations of 30\,\%, 50\,\% and 60\,\% are used. The cross section TEM image of the 60\,\% Co:ZnO/Py sample as well as the electron diffraction pattern of the Co:ZnO layer (b) and a magnification on the interface between Co:ZnO and Py (c) are shown.}
	\label{figure1}
\end{figure}

Figure\,\ref{figure1}(a) displays the four different types of samples: Co:ZnO layers, with Co concentrations of 30\,\%, 50\,\% and 60\,\%, are grown with a nominal thickness of 100\,nm and Py with 10\,nm. To prevent surface oxidation a capping layer of 5\,nm Al is used. For single 60\,\% Co:ZnO films the vertical-exchange bias effect was largest compared to lower Co concentrations. Therefore, for 60\,\% Co:ZnO samples with an additional Al layer as spacer between Co:ZnO and Py have been fabricated. The thickness of the Al spacer (1\,nm, 1.5\,nm and 2\,nm) is in a range where the Al is reported not to suppress spin pumping effects itself \cite{KSA13}. 

\subsection*{TEM}

To get information about the interface between Py and Co:ZnO high resolution cross section transmission electron microscopy (TEM) was done. In Fig.\,\ref{figure1}(b) the cross section TEM image of 60\,\% Co:ZnO/Py with the electron diffraction pattern of the Co:ZnO is shown. A magnification of the interface between Co:ZnO and Py is shown in Fig.\,\ref{figure1}(c). From XRD measurements \cite{NHL16} it is obvious that the quality of the wurtzite crystal slightly decreases for higher Co doping in ZnO. A similar behavior is observed in TEM cross section images. While 35\,\% Co:ZnO shows the typical only slightly misoriented columnar grain growth \cite{NHL16} it is obvious from Fig.\,\ref{figure1}(b) that the crystalline nanocolumns are less well ordered for 60\,\% Co:ZnO. Although the electron diffraction pattern confirms a well ordered wurtzite structure, the misorientation of lattice plains is stronger than for 35\,\% Co:ZnO \cite{NHL16}, even resulting in faint Moir{\'e} fringes which stem from tilted lattice plains along the electron path. This corroborates previous findings of $\omega$-rocking curves in XRD \cite{NHL16, BHN18} where the increase in the full width at half maximum also evidences a higher tilting of the crystallites, i.e. an increased mosaicity. The interface to the Py layer is smooth, although it is not completely free of dislocations. Also the interface seems to be rather abrupt within one atomic layer, i.e. free of intermixing. A similar behavior is found for the interface between 50\,\% Co:ZnO and Py (not shown).

\subsection*{XAS and XMCD}

\begin{figure}[h]
	\centering
		\includegraphics[width=0.45\textwidth]{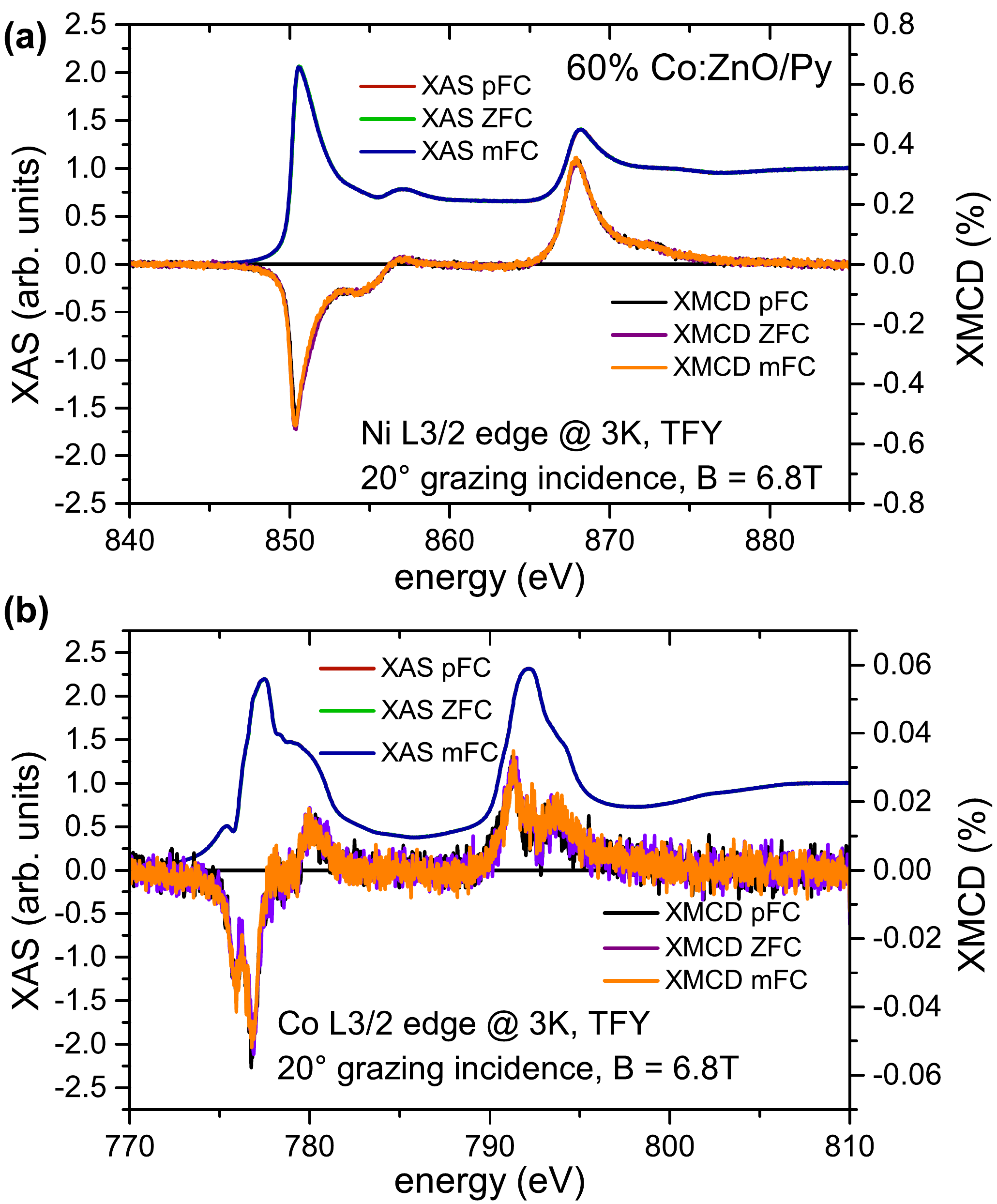}
		\vspace{-0.2cm}
	\caption{In (a) the XMCD at the Ni L$_\text{3/2}$ edges after pFC, mFC and ZFC for 60\,\% Co:ZnO/Py are shown. (b) shows the same for the Co L$_\text{3/2}$ edges.}
	\label{figure2}
\end{figure}

Figure\,\ref{figure2} shows XAS and XMCD spectra recorded at 3\,K and a magnetic field of 6.8\,T at the Ni L$_\text{3/2}$ and Co L$_\text{3/2}$ edges of 60\,\% Co:ZnO/Py after pFC, mFC or ZFC. For all three cooling conditions the Ni L$_\text{3/2}$ edges (Fig.\,\ref{figure2}(a)) show a metallic character of the Ni XAS without any additional fine structure characteristics for NiO and thus no sign of oxidation of the Py. Further, no differences in the XAS or the XMCD of the Ni edges of different cooling conditions are found. The same is observed for the Fe L$_\text{3/2}$ edges, however, they are affected greatly by self-absorption processes in total fluorescence yield (not shown). 

The Co L$_\text{3/2}$ edges in Fig.\,\ref{figure2}(b) are also greatly affected by the self absorption of the total fluorescence yield, since it is buried below 10\,nm of Py and 5\,nm of Al. In contrast to Ni the XAS and XMCD at the Co L$_\text{3/2}$ edges (Fig.\,\ref{figure2}(b)) are not metallic and evidence the incorporation of Co as Co$^{2+}$ in the wurtzite structure of ZnO \cite{NHL16, BHN18}. The overall intensity of the Co XMCD is strongly reduced indicating a small magnetic moment per Co atom well below metallic Co. This small effective Co moment in 60\,\% Co:ZnO can be understood by the degree of antiferromagnetic compensation that increases with higher Co doping concentrations \cite{NHL16}. Furthermore, no indications of metallic Co precipitates are visible in the XAS and XMCD of the heterostructure as it would be expected for a strong intermixing at the interface to the Py.

No changes between the pFC, mFC and ZFC measurements are visible also for the Co edges either in XAS or XMCD indicating that the spin system of the Co dopants is not altered in the exchange bias state. This corroborates measurements conducted at the Co K-edge \cite{BHN18}. After field cooling the XMCD at the Co main absorption increased compared to the ZFC conditions. At the Co K-edge the main absorption stems from the orbital moment. The spin system is only measured indirectly at the pre-edge feature which remained unaffected by the cooling field conditions. The data of K- and L-edges combined evidences that the imprinted magnetization after field cooling is composed predominantly of orbital moment, which is in good agreement with other EB systems \cite{SSD10, ASB16}

%The vertical shift of single Co:ZnO layers could be attributed to an increased orbital moment of pinned uncompensated AFM moments \cite{BHN18}. The XMCD at the Co K-edge showed a stronger main absorption after field-cooling 50\,\% Co:ZnO and 60\,\% Co:ZnO, which is connected to the orbital moment, while the pre-edge, connected to the spin system was nearly unaffected. Also the Co L$_\text{3/2}$ edges of the field-cooled heterostructures does not exhibit a stronger XMCD for 60\,\% Co:ZnO/Py corroborating the information from the K-edge, that the vertical shift in Co:ZnO is caused by an increased orbital moment. 

\subsection*{SQUID}
\begin{figure}[h]
	\centering
		\includegraphics[width=0.45\textwidth]{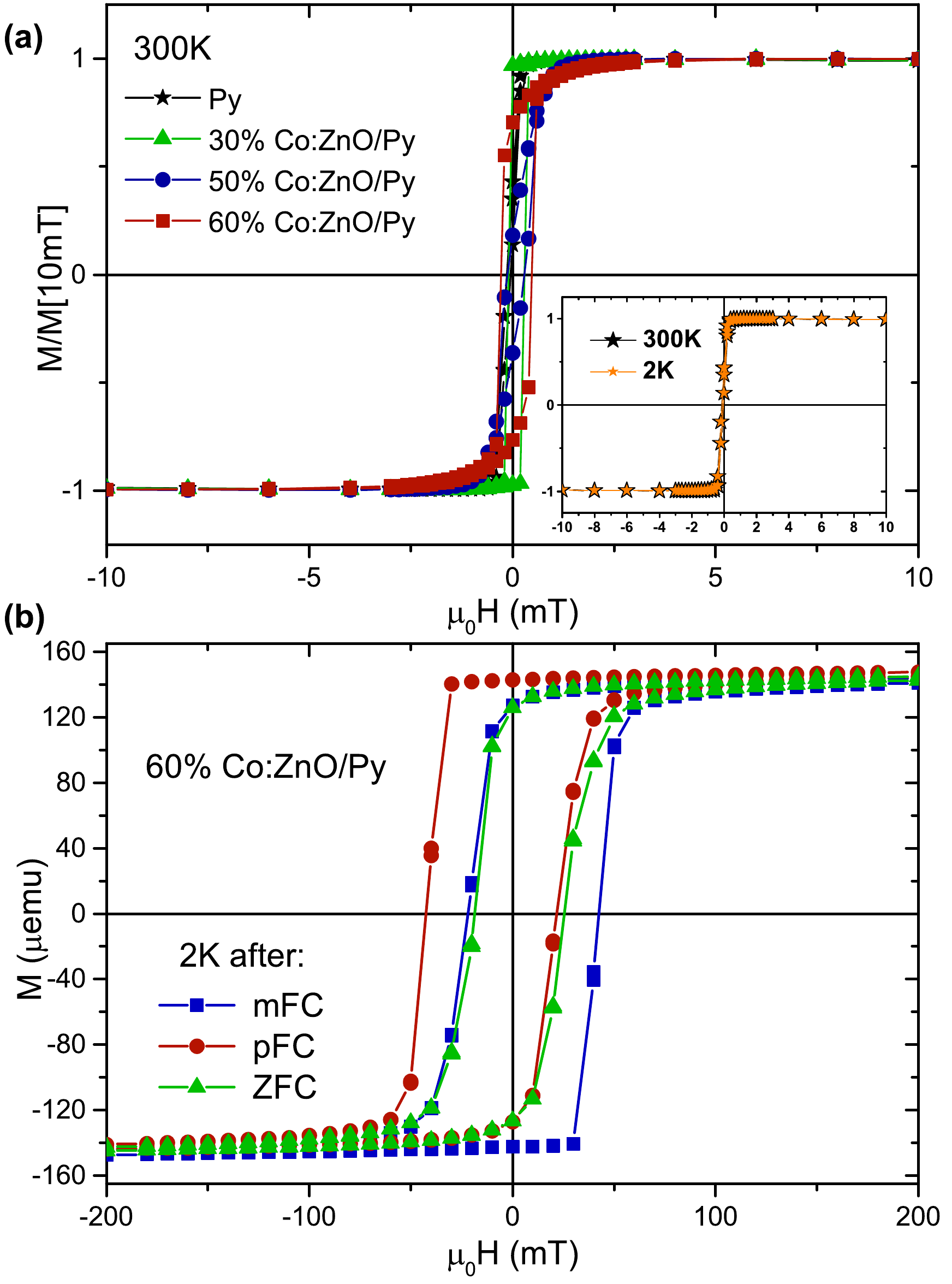}
		\vspace{-0.2cm}
	\caption{At 300\,K the $M(H)$ curves of the single Py film almost overlaps with the $M(H)$ curves of the heterostructures with all three Co:ZnO concentrations (a). In the inset it can be seen that there is no difference in coercive field for Py at 300\,K and 2\,K. Measuring the 60\,\% Co:ZnO/Py heterostructure after plus, minus and zero field cooling, horizontal and vertical exchange bias shifts are visible, as well as an increase in the coercive field (b).}
	\label{figure3}
\end{figure}

The static coupling in the heterostructures was investigated by integral SQUID magnetometry. Measurements done at 300\,K, as shown in Fig.\,\ref{figure3}(a), do not reveal a significant influence of the Co:ZnO on the $M(H)$ curve of Py. Just a slight increase in coercive field from 0.1\,mT to 0.4\,mT is determined. Some of the $M(H)$ curves in Fig.\,\ref{figure3}(a) are more rounded than the others. This can be attributed to slight variations in the aspect ratio of the SQUID pieces and thus variations in the shape anisotropy. The inset of Fig.\,\ref{figure3}(a) shows the hysteresis of the single Py film at 300\,K and 2\,K, where no difference in coercivity is visible. Please note that up to now measurements were conducted only in a field range of $\pm$10\,mT and directly after a magnet reset. This is done to avoid influences of the offset field of the SQUID \cite{BHH18}. At low temperatures, to determine the full influence of Co:ZnO, high fields need to be applied, as it has been shown in \cite{BHN19}. Therefore, coercive fields obtained from low temperature measurements are corrected by the known offset field of 1.5\,mT of the SQUID \cite{BHH18}.

 Since the paramagnetic signal of Co:ZnO is close to the detection limit of the SQUID and thus, orders of magnitude lower than the Py signal it has no influence on the room temperature $M(H)$ curve. However, with an additional Co:ZnO layer a broadening of the hysteresis, a horizontal and a small vertical shift are measured at 2\,K as can be seen exemplary for 60\,\% Co:ZnO/Py in Fig.\,\ref{figure3}(b). Similar to single Co:ZnO films where an opening of the $M(H)$ curve is already visible in ZFC measurements \cite{HNS16, NHL16, BHN18,BHN19} also in the heterostructure no field cooling is needed to increase the coercive field.

 \begin{figure}[h]
	\centering
		\includegraphics[width=0.45\textwidth]{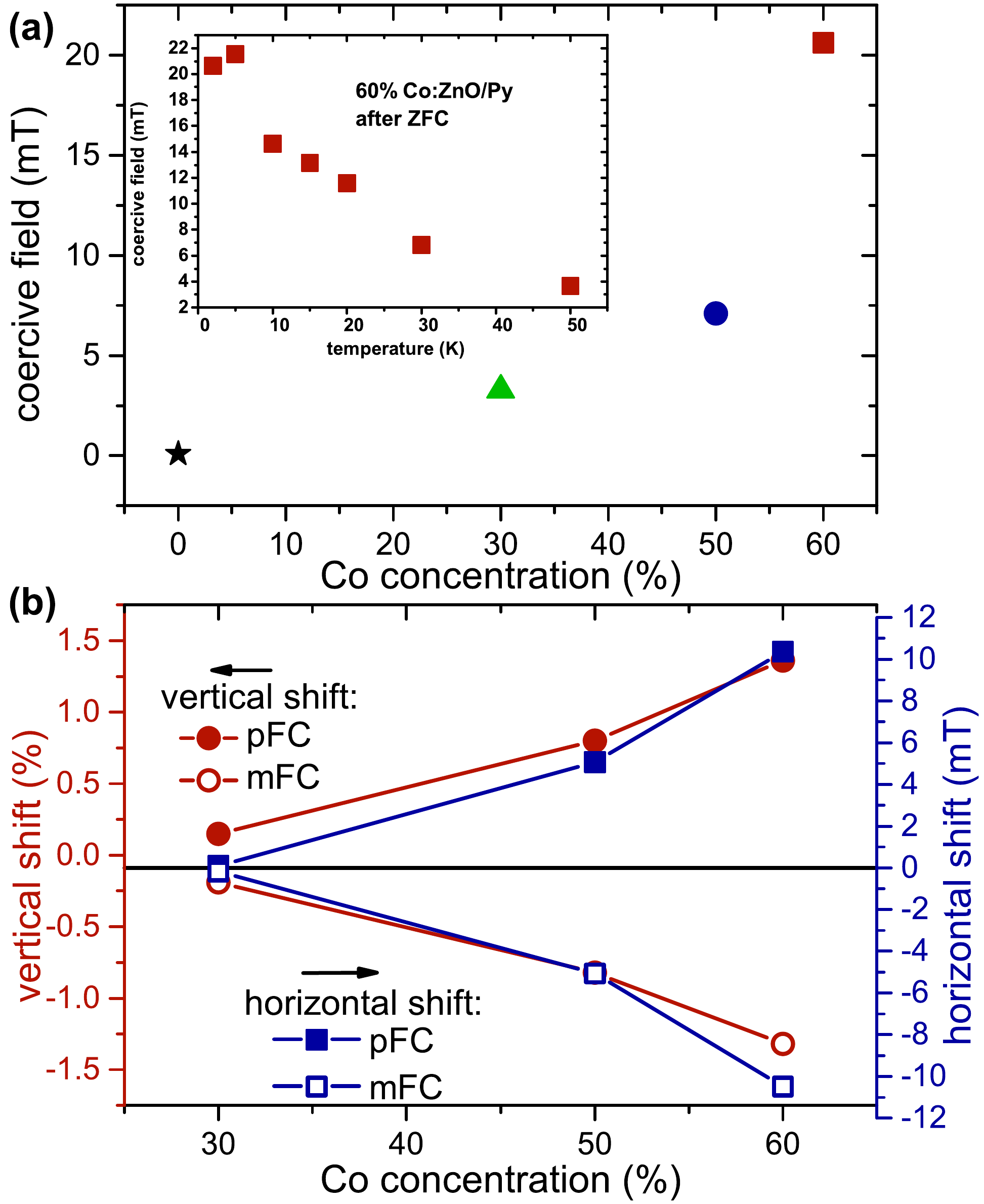}
		\vspace{-0.2cm}
	\caption{(a) At 2\,K the coercivity increases with Co concentration in the heterostructure. In the inset the temperature dependence of the coercivity of the 60\,\% Co:ZnO/Py heterostructure is given. (b) The vertical shift (circles) and the horizontal shift (squares) depend on the Co concentration. Both shifts reverse the direction when the measurement is changed from pFC to mFC.}
	\label{figure4}
\end{figure}

Earlier works \cite{NHL16, HNS16} demonstrated that the hysteresis opening and vertical shift in Co:ZnO are strongly dependent on the Co concentration and increase with increasing Co doping level. Furthermore, the EB effects are observed in the in-plane and out-of-plane direction, with a greater vertical shift in the plane. Therefore, the heterostructers with Py are measured with the magnetic field in in-plane direction. Figure\,\ref{figure4}(a) provides an overview of the coercive field after ZFC for the different Co concentrations. The coercive field increases from 0.1\,mT for single Py to 20.6\,mT for 60\,\% Co:ZnO/Py. Additionally, in the inset the temperature dependence of the coercive field of the 60\,\% Co:ZnO/Py heterostructure is shown, since it shows the strongest increase in coercive field. From the 20.6\,mT at 2\,K it first increases slightly when warming up to 5\,K. That the maximum coercivity is not at 2\,K is in good agreement with measurements at single 60\,\% Co:ZnO films where a maximum hysteresis opening at 7\,K was determined \cite{BHN19}. Afterwards the coercive field decreases. At the N\'{e}el temperature of 20\,K a coercive field of 11.6\,mT is measured. Above T$_\text{N}$ it decreases even further but the coercivity is still 3.65\,mT at 50\,K. A coupling above T$_\text{N}$ could stem from long range magnetic ordered structures in Co:ZnO where first indications are visible already in single Co:ZnO films \cite{NHL16}. However, for single layers they are barely detectable with the SQUID.  

The vertical (circles) and horizontal (squares) hysteresis shifts after pFC and mFC are shown in Fig.\,\ref{figure4}(b) for the Py samples with Co:ZnO layers. Similar to single Co:ZnO films the vertical shift increases with rising Co concentration. The shift is given in percent of the magnetization at 5\,T to compensate for different sample sizes. Due to the overall higher magnetization at 5\,T in combination with Py this percentage for the heterostructures is lower than the vertical shift for single Co:ZnO films. With increasing Co concentration the degree of antiferromagnetic compensation increases \cite{NHL16, BHN19}, which in turn should lead to a stronger EB coupling. This can be seen in the horizontal shift and thus EB field which is strongest for 60\,\% Co:ZnO/Py and nearly gone for 30\,\% Co:ZnO/Py. For both kinds of shift the pFC and mFC measurements behave similar, except the change of direction of the shifts. 

\subsection*{Multifrequency FMR}

\begin{figure}[h]
	\centering
		\includegraphics[width=0.45\textwidth]{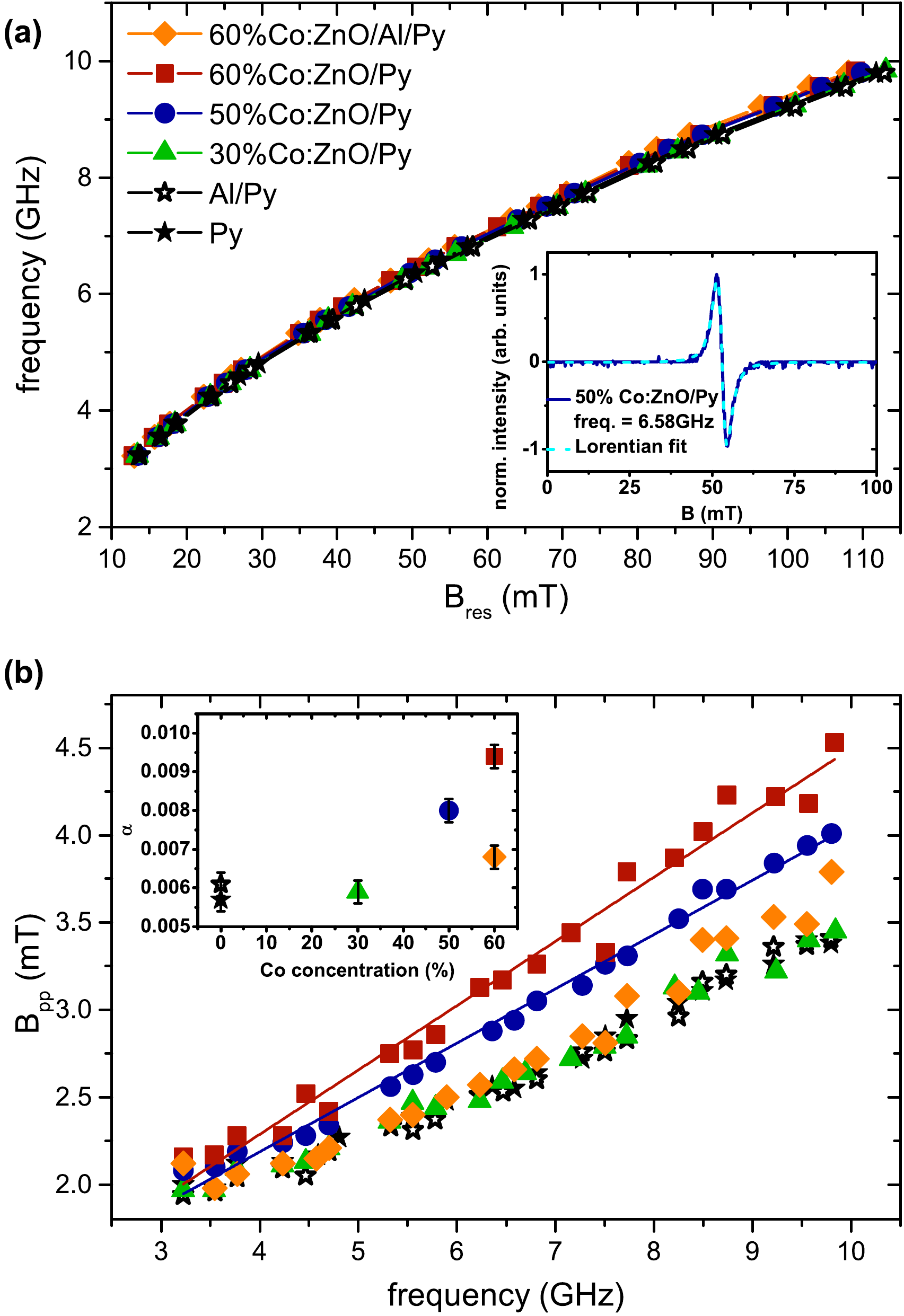}
		\vspace{-0.2cm}
	\caption{The resonance fields determined at room temperature with multifrequency FMR are seen in (a). In the inset an exemplary FMR spectrum for of 50\,\% Co:ZnO/Py at 6.58\,GHz is shown with the corresponding Lorentian fit. For the linewidth (b) and the associated damping parameter $\alpha$ (inset) an increase is visible for the heterostructures with higher Co concentration in the Co:ZnO. The lines are linear fits to the data.
	}
	\label{figure5}
\end{figure}

The dynamic coupling between the two layers has been investigated by multifrequency FMR measured at room temperature. The frequency dependence of the resonance position between 3\,GHz and 10\,GHz of the heterostructures is shown in Fig.\,\ref{figure5}(a). The resonance position of Py yields no change regardless of the Co concentration in the Co:ZnO layer or its complete absence. Also in 2\,nm Al/Py and 60\,\% Co:ZnO/2\,nm Al/Py the resonance position stays unchanged. The resonance position of a thin film is given by Kittel formula \cite{Kittel48}:

\begin{equation}
f = \frac{\gamma}{2 \pi} \sqrt{B_\text{res} \left(B_\text{res} + \mu_\text{0} M \right)}
\label{eq1}
\end{equation}

with the gyromagnetic ratio $\gamma = \frac{g \mu_\text{B}}{\hbar}$ and magnetization $M$. However, any additional anisotropy adds to $B_\text{res}$ and therefore alters eq.\,(\ref{eq1}) \cite{Kittel48}. The fact that all samples show the identical frequency dependence of the resonance position evidences that neither the gyromagnetic ratio $\gamma$ and thus the Py $g$-factor are influenced nor any additional anisotropy $B_\text{Aniso}$ is introduced by the Co:ZnO. By fitting the frequency dependence of the resonance position using the Kittel equation with the g-factor of 2.11 \cite{MAM02} all the samples are in the range of (700$\pm$15)\,kA/m, which within error bars is in good agreement with the saturation magnetization of (670$\pm$50)\,kA/m determined from SQUID.

Even though the Co:ZnO layer does not influence the resonance position of the FMR measurement the heterostructures exhibit an increase in linewidth. This corresponds to a change of the damping in the system. The frequency dependence of the linewidth can be used to separate the inhomogeneous from the homogeneous (Gilbert like) contributions, from which the Gilbert damping parameter $\alpha$ can be determined.

\begin{equation}
\Delta B = \Delta B_\text{hom} + \Delta B_\text{inhom}
\end{equation}

where
\begin{equation}
\Delta B_\text{hom} = \frac{4 \pi \alpha}{\gamma} f
\end{equation}
No difference in linewidth between Al/Py (open stars) and Py (full stars) is found, as can be seen in Fig.\,\ref{figure4}(b) where the peak to peak linewidth B$_\text{pp}$ is plotted over the measured frequency range for all the heterostructures. While the heterostructure with 30\,\% Co:ZnO/Py (green triangles) lies atop the single Py and the Al/Py film, the linewidth increases stronger with frequency for 50\,\% Co:ZnO/Py (blue circles). The broadest FMR lines are measured for the 60\,\% Co:ZnO/Py heterostructure (red sqaures). 

Using the Py $g$-factor of 2.11 \cite{MAM02}, $\alpha$ can be calculated from the slopes of the frequency dependence extracted from the linewidths seen in Fig.\,\ref{figure5}(b): the resulting $\alpha$ are shown in the inset. For the single Py layer $\alpha_\text{Py}$ = (5.7$\pm$0.3)$\times$10$^\text{-3}$ which compares well to previously reported values \cite{TBB02}. This increases to $\alpha_\text{50}$ = (8.0$\pm$0.3)$\times$10$^\text{-3}$ for 50\,\% Co:ZnO/Py and even $\alpha_\text{60}$ = (9.4$\pm$0.3)$\times$10$^\text{-3}$ for 60\,\% Co:ZnO/Py. So the damping increases by a factor of 1.64 resulting in a spin pumping contribution $\Delta \alpha$ = (3.7$\pm$0.5$)\times$10$^\text{-3}$ that stems from the angular momentum transfer at the interface of Py and Co:ZnO. By insertion of a 2\,nm Al spacer layer $\Delta \alpha$ reduces to (0.8$\pm$0.5)$\times$10$^\text{-3}$. 

\subsection*{Dependence on the Al spacer thickness}

\begin{figure}[h]
	\centering
		\includegraphics[width=0.45\textwidth]{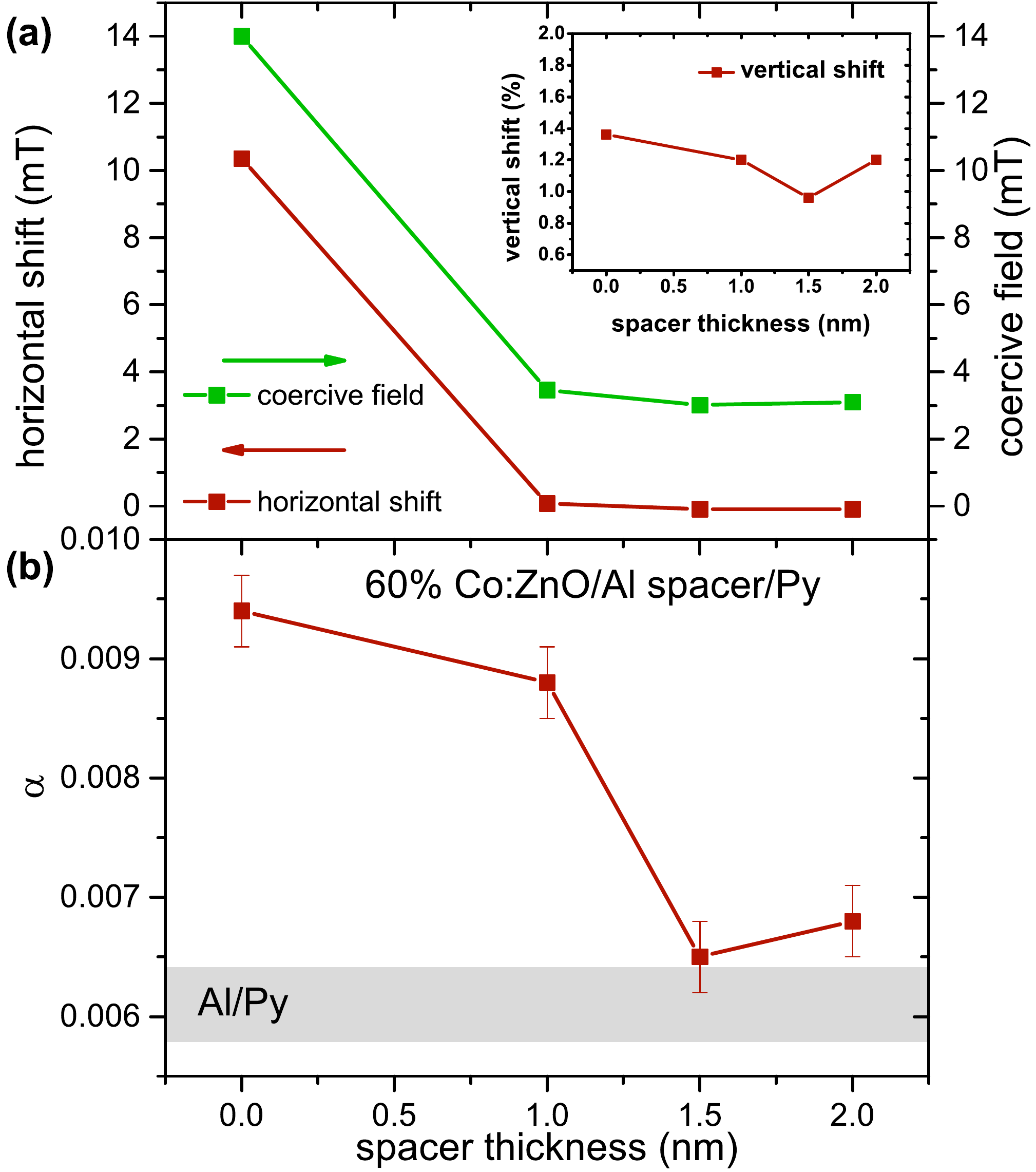}
		\vspace{-0.2cm}
	\caption{When an Al spacer is inserted between the Py and the Co:ZnO layer horizontal shift and coercive field show a strong decrease already at 1\,nm spacer thickness (a) while the vertical shift (inset) is not dependent on the spacer thickness. (b) shows the effect of the Al spacer on the Gilbert damping parameter $\alpha$, which also decreases if the spacer gets thicker than 1\,nm. As shaded region the Gilbert damping parameter of a Al/Py film is indicated within error bars.}
	\label{figure6}
\end{figure}

To obtain information about the lengthscale of the static and dynamic coupling, heterostructures with Al spacer layers of different thickness (1\,nm, 1.5\,nm and 2\,nm thick) between Py and the material beneath (sapphire substrate or 60\,\% Co:ZnO) were fabricated. Without a Co:ZnO layer the spacer underlying the Py layer does not exhibit any changes in either SQUID (not shown) or FMR (see Fig\,\ref{figure5} (a) and (b)). The results obtained for the 60\,\% Co:ZnO/Al/Py heterostructure for the coercive field, vertical and horizontal shift extracted from $M(H)$ curves are shown in Fig.\,\ref{figure6}(a), whereas the damping parameter $\alpha$ from room temperature multifrequency FMR measurements, analogues to Fig.\,\ref{figure5}(b), are depicted in Fig.\,\ref{figure6}(b). 

The horizontal shift and the increased coercive field are caused by the coupling of FM and AFM moments in range of a few \AA ngstrom to the interface \cite{FKR07, BSR08, BSG08}. Therefore, both effects show a similar decrease by the insertion of an Al spacer. While the horizontal shift and coercive field are reduced significantly already at a spacer thickness of 1\,nm, the vertical shift (inset of Fig.\,\ref{figure6}(a)) is nearly independent of the Al spacer. Comparing with the XMCD spectra of Fig.\,\ref{figure2} it can be concluded that the vertical shift in the uncompensated AFM/FM system Co:ZnO/Py stems solely from the increased orbital moment of pinned uncompensated moments in Co:ZnO and is independent of the FM moments at the interface. Furthermore, the FM moments do not exhibit any vertical shift and the exchange between the two layers only results in the horizontal shift.

For the FMR measurments after inserting an Al spacer no effect on the resonance position is found, as was shown already in Fig.\,\ref{figure5}(a). For a 1\,nm thick Al spacer the damping results in $\alpha$ = (8.8$\pm$0.3)$\times$10$^\text{-3}$, which gives a $\Delta \alpha$ = (3.1$\pm$0.5)$\times$10$^\text{-3}$. This is only a slight decrease compared to the sample without Al spacer. By increasing the spacer thickness $\alpha$ reduces to values just above the damping obtained for pure Py or Al/Py, shown as shaded region in Fig.\,\ref{figure6}(b). The 1\,nm thick Al layer is thick enough to suppress intermixing between the Co:ZnO and the Py layer as can be seen in Fig.\,\ref{figure1}(b). Together with the unchanged behavior of Al/Py without Co:ZnO damping effects due to intermixing between Al and Py can be excluded. Also, a change in two magnon scattering can be ruled out, since it would account for non-linear effects on the linewidth and contribute to $\Delta B_\text{inhom}$ \cite{Bab11}. Therefore, the increase in Gilbert damping can be attributed to a dynamic coupling, e.g. spin pumping from Py into Co:ZnO. Furthermore, the dynamic coupling mechanism is extends over a longer range than the static coupling. With 1\,nm spacer the dynamic coupling is only slightly reduced whereas the static coupling is already completely suppressed. 

%So similar as for the static coupling also the dynamic coupling nearly vanishes by separating Py and Co:ZnO by more than 1\,nm of Al showing that the majority of magnetic coupling between Py and Co:ZnO is confined near the interface.

\subsection*{Temperature dependent FMR}
\begin{figure*}[ht]
	\centering
		\includegraphics[width=0.84\textwidth]{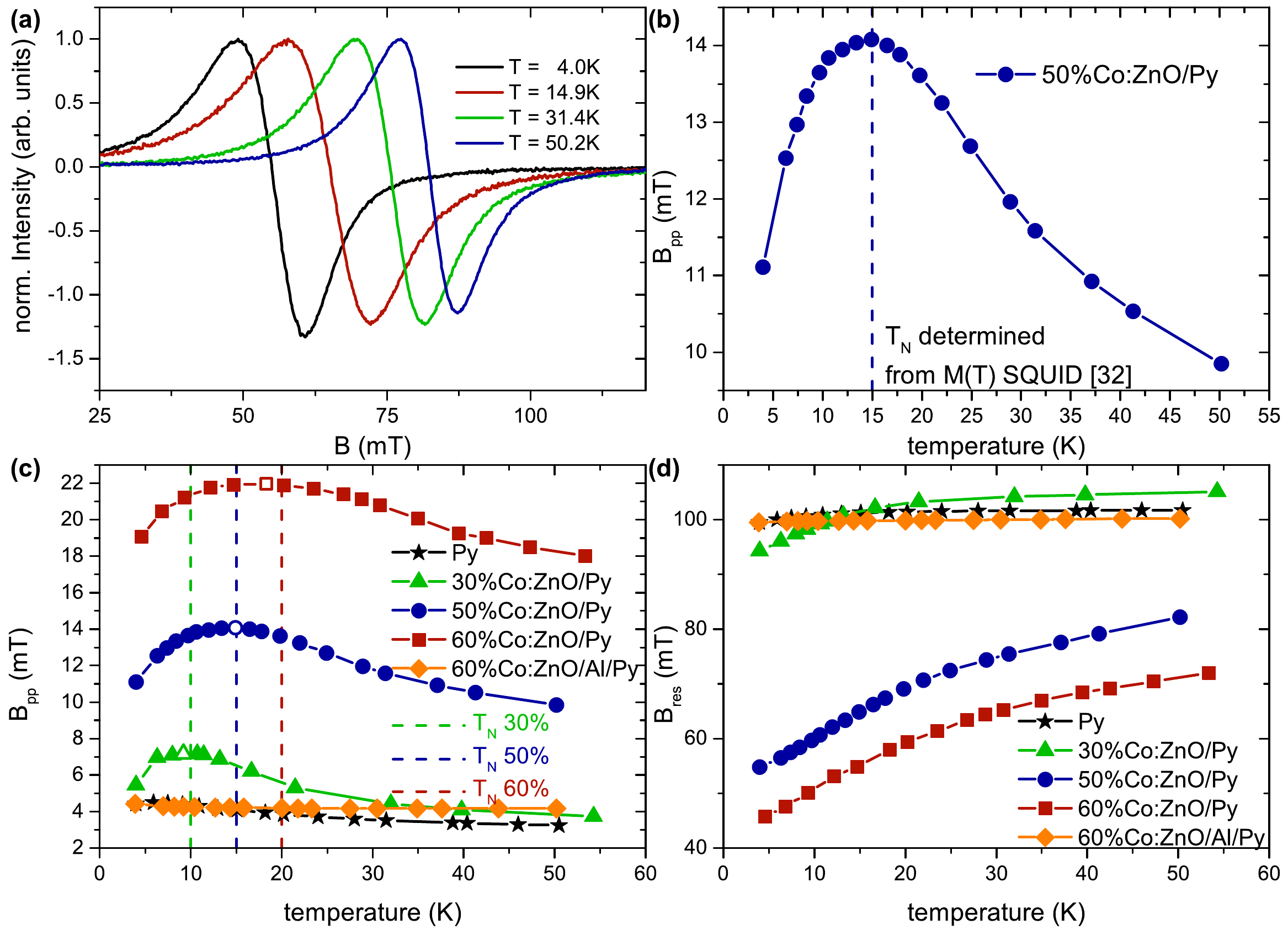}
		\vspace{-0.2cm}
	\caption{By decreasing the temperature the resonance position of 50\,\% Co:ZnO/Py shifts to lower resonance fields (a) and the linewidth increases, showing a maxium at the T$_\text{N}$ (b). A similar behavior is observed for the heterostructures with 30\,\% and 60\,\% Co doping while a single Py film does not exhibit a maximum when cooling (c). The maximum is marked as open symbol in the temperature dependence, while the T$_\text{N}$ determined from $M(T)$ \cite{NHL16} are shown as dashed lines. Furthermore, the resonance position of the heterostructures with Co:ZnO shifts at low temperatures (d).}
	\label{figure7}
\end{figure*} 

In vicinity to the magnetic phase transition temperature the spin pumping efficiency should be at a maximum \cite{OAS14, FOA16}. Therefore, the samples are measured inside a resonator based FMR setup, as a function of temperature. During the cooldown no magnetic field is applied and the results shown in Fig.\,\ref{figure7} are ZFC measurements. For 50\,\% Co:ZnO/Py the resonance positions shifts of Py to lower magnetic fields as the temperature decreases as can be seen in Fig.\,\ref{figure7}(a). Not only the resonance position is shifting, but also the linewidth is changing with temperature as shown in Fig.\,\ref{figure7}(b). The linewidth has a maximum at a temperature of 15\,K which corresponds well to T$_\text{N}$ determined by $M(T)$ SQUID measurements for a 50\,\% Co:ZnO layer \cite{NHL16}. This maxium of the linewidth in the vicinity of T$_\text{N}$ is also observed for 60\,\% Co:ZnO/Py and even 30\,\% Co:ZnO/Py, as shown in Fig.\,\ref{figure7}(c). The measured maximum of 30\,\% Co:ZnO/Py and 60\,\% Co:ZnO/Py are at 10.7\,K, 19.7\,K respectively and are marked with an open symbol in Fig.\,\ref{figure7}(c). For comparison the N{\'e}el temperatures determined from $M(T)$ measurements \cite{NHL16} are plotted as dashed line. Py on the other hand shows only a slight increase in linewidth with decreasing temperature. The observed effects at low temperatures vanish for the 60\,\% Co:ZnO/2\,nm Al/Py heterostructure.

Figure\,\ref{figure7}(d) shows the temperature dependence of the resonance field for all samples. For Py $B_\text{res}$ only decreases slightly whereas for 50\,\% and 60\,\% Co:ZnO a strong shift of $B_\text{res}$ can be observed. This shift evidences a magnetic coupling between the Py and the Co:ZnO layer. Even in the heterostructure with 30\,\% Co:ZnO/Py a clear decrease in resonance position below 10\,K (the previously determined T$_\text{N}$ \cite{NHL16}) is visible. This shift of the resonance position is only observed at low temperatures. At room temperature no shift of the resonance position at 9.5\,GHz has been observed as shown in Fig.\,\ref{figure5}(a). From the low-temperature behavior of the single Py layer and eq.\,\ref{eq1} it is obvious that the gyromagnetic ratio is not changing strongly with temperature, therefore shift of the resonance position in the heterostructure can be attributed to a change in anisotropy. From the SQUID measurements at 2\,K, see Fig.\,\ref{figure3}(b) and Fig.\,\ref{figure4}(b) EB between the two layers has been determined, which acts as additional anisotropy \cite{MB56} and therefore causes the shift of the resonance position.
Both the shift of the resonance position and the maximum in FMR linewidth vanish if the Py is separated from 60\,\% Co:ZnO by a 2\,nm Al spacer layer. So, also at low temperatures the static EB coupling and the dynamic coupling can be suppressed by an Al spacer layer.

 %A stronger decrease is observed for 50\,\% Co:ZnO/Py and 60\,\% Co:ZnO/Py lasting also to temperatures well above their respective T$_\text{N}$.

$M(T)$ measurements indicated a more robust long-range magnetic order in 60\,\% Co:ZnO by a weak separation of the field heated and ZFC curves lasting up to 200\,K \cite{NHL16}. Additionally, the coercive field measurements on the 60\,\% Co:ZnO/Py hetersotructure revealed a weak coupling above T$_\text{N}$. However, this has not been observed for lower Co concentrations. In the heterostructure with 30\,\% Co:ZnO the FMR resonance position and linewidth return quickly to the room temperature value for temperatures above the T$_\text{N}$ of 10\,K. For both 50\,\% Co:ZnO/Py and 60\,\% Co:ZnO/Py the resonance positions are still decreased and the linewidths are increased above their respective N{\'e}el temperatures and are only slowly approaching the room temperature value. In the 60\,\% Co:ZnO/Py heterostructure measurements between 100\,K and 200\,K revealed that a reduced EB is still present. It is known for the blocking temperatures of superparamagnetic structures that in FMR a higher blocking temperature compared to SQUID is obtained due to much shorter probing times in FMR of the order of nanoseconds compared to seconds in SQUID \cite{ALF05}. Hence, large dopant configurations in Co:ZnO still appear to be blocked blocked on timescales of the FMR whereas they already appear unblocked on timescales of the SQUID measurements.

\section*{V. Conclusion}

The static and dynamic magnetic coupling of Co:ZnO, which is weakly paramagnetic at room temperature and an uncompensated AFM at low temperatures, with ferromagnetic Py was investigated by means of SQUID magnetometry and FMR. At room temperature no static interaction is observed in the $M(H)$ curves. After cooling to 2\,K an EB between the two layers is found resulting in an increase of coercive field and a horizontal shift. Additionally, a vertical shift is present caused by the uncompensated moments in the Co:ZnO. While this vertical shift is nearly unaffected by the insertion of an Al spacer layer between Co:ZnO and Py the EB vanishes already at a spacer thickness of 1\,nm. 

The FMR measurements at room temperature reveal an increase of the Gilbert damping parameter for 50\,\% Co:ZnO/Py and 60\,\% Co:ZnO/Py, whereas 30\,\% Co:ZnO/Py is in the range of an individual Py film. At room temperature the resonance position is not affected for all the heterostructures. For the 60\,\% Co doped sample $\Delta \alpha$ = 3.7$\times$10$^\text{-3}$, which is equivalent to an increase by a factor of 1.64. In contrast to the static magnetic coupling effects, an increased linewidth is still observed in the heterostructure containing a 1\,nm Al spacer layer. 

At lower temperatures the resonance position shifts of the heterostructures to lower resonance fields, due to the additional EB anisotropy. The temperature dependence of the linewidth shows a maximum at temperatures, which by comparison with $M(T)$ measurements correspond well to T$_\text{N}$ of single Co:ZnO layers and thus corroborate the increase of the damping parameter and thus spin pumping efficiency in vicinity to the magnetic phase transition. Furthermore, the shift of the resonance position has been observed at temperatures well above T$_\text{N}$ for 50\,\% Co:ZnO/Py and 60\,\% Co:ZnO/Py. Up to now only indications for a long range AFM order in 60\,\% Co:ZnO/Py had been found by static $M(T)$ measurements. The dynamic coupling, however, is sensitive to those interactions due to the higher time resolution in FMR resulting in a shift of the resonance position above the T$_\text{N}$ determined from $M(T)$ SQUID.  
   
\section*{Acknowledgment}

The authors gratefully acknowledge funding by the Austrian Science Fund (FWF) - Project No. P26164-N20 and Project No. ORD49-VO. All the measured raw data can be found in the repository at http://doi.org/10.17616/R3C78N. The x-ray absorption measurements were performed on the EPFL/PSI X-Treme beamline at the Swiss Light Source, Paul Scherrer Institut, Villigen, Switzerland. Furthermore, the authors thank Dr. W. Ginzinger for the TEM sample preparation and measurements.

\end{document}